\documentclass[a4paper,reprint,aps,prl,superscriptaddress,showpacs]{revtex4-1}

	\usepackage[utf8]{inputenc}
	\usepackage[T1]{fontenc}
	\usepackage[english]{babel}
	\usepackage{graphicx}
	\usepackage{amsfonts}
	\usepackage{amsmath}
	\usepackage{amssymb}
	\usepackage{subfigure}
	\usepackage[colorlinks=true,citecolor=blue,linkcolor=blue,urlcolor=blue]{hyperref}

\begin{document}

\title{Plasma-driven ultrashort bunch diagnostics}

\author{I. Dornmair}
	\affiliation{Center for Free-Electron Laser Science \& Department of Physics\\
University of Hamburg, Luruper Chaussee 149, 22761 Hamburg, Germany}
	
\author{C. B. Schroeder}
    \affiliation{Lawrence Berkeley National Laboratory, Berkeley, California 94720, USA}
    
\author{K. Floettmann}
	\affiliation{DESY, 22607 Hamburg, Germany}
    
\author{B. Marchetti}
   	\affiliation{DESY, 22607 Hamburg, Germany}
   	
\author{A. R. Maier}
	\email{andreas.maier@desy.de}
	\affiliation{Center for Free-Electron Laser Science \& Department of Physics\\
University of Hamburg, Luruper Chaussee 149, 22761 Hamburg, Germany}

\begin{abstract}
Ultrashort electron bunches are crucial for an increasing number of applications, however, diagnosing their longitudinal phase space remains a challenge. We propose a new method that harnesses the strong electric fields present in a laser driven plasma wakefield. By transversely displacing driver laser and witness bunch, a streaking field is applied to the bunch. This field maps the time information to a transverse momentum change and, consequently, to a change of transverse position. We illustrate our method with simulations where we achieve a time resolution in the attosecond range.
\end{abstract}


\date{\today}

\maketitle

The production of ultrashort electron bunches is crucial for many applications such as ultrafast electron diffraction \cite{zewail2006} or short pulse free-electron laser (FEL) operation \cite{neutze2000}, as the bunch length and consequently X-ray radiation pulse length determines the achievable time resolution and can circumvent limitations from sample damage. Complementary to RF-based accelerators \cite{assmannsinbad2014,zeitler2015}, several novel acceleration techniques promise to create electron bunches around or below the few femtosecond (fs) length, such as laser-plasma accelerators (LPA) \cite{esarey2009,buck2011,lundh2011} or dielectric structures \cite{england2014}. However, established methods to measure the longitudinal phase space suffer from severe limitations. The temporal resolution of electro-optic monitors is limited to 50\,fs \cite{berden2007,pompili2014}. Coherent transition radiation measurements are able to diagnose ultrashort bunch lengths, yet this is an indirect method that, due to the lack of phase information, provides no unique solution for the reconstructed bunch shape \cite{lundh2011,bajlekov2013}.

Typically, the longitudinal phase space is measured with a transverse deflecting structure (TDS) \cite{emma2000,roehrs2009}. Similar to a streak camera, a TDS uses an RF cavity to imprint a linear transverse momentum change along the bunch. In a subsequent drift or imaging beam optic this momentum change transfers the longitudinal information into a transverse position. In order to achieve resolutions down to 1\,fs \cite{behrens2009,dolgashev2012,orlandi2013,behrens2014} these cavities are typically several meters long. This limits their applicability to compact accelerators like LPA or to bunches that undergo a short temporal focus \cite{Lu2015,hada2012}, and increases detrimental effects such as accumulated energy chirp or induced beam offset \cite{floettmann2014_tds}.

Here, we propose a novel technique that applies a laser-driven wakefield to streak the electron bunch. We achieve resolutions well below \mbox{1 fs}, owing to the high field amplitude and frequency of the plasma wake. Precise knowledge of the electron phase space could then be used to optimize a compression scheme, or, in the case of LPA, to gain information and control over the injection mechanism.

In the following, we review the theoretical treatment of transverse deflecting structures \cite{behrens2009,floettmann2014_tds} and apply it to laser-driven wakefields. We then discuss two different experimental setups: a collinear geometry with co-propagating electron bunch and laser, and a setup with an angle between laser and electron beam path. We illustrate our method with particle in cell (PIC) simulations, and discuss limitations on the resolution from beam loading and energy spread.

The transverse momentum change imprinted by a TDS can be written as
\begin{equation}
p_y(\xi) = \frac{eV}{c}\sin{(k\xi+\Psi_0)},
\end{equation}
where $V$ is the effective voltage given by the integral of the transverse electric and magnetic field components over the structure length, $k$ is the wavenumber, $\xi$ the longitudinal internal bunch coordinate and $\Psi_0$ defines the phase of the bunch with respect to the field.
Ideally, the bunch is situated at the zero-crossing of the field ($\Psi_0=0$), so that the bunch centroid remains unaffected and the imprinted deflection is nearly linear along the bunch. The linearized angular deflection of the beam per unit length is
\begin{equation}
S = \frac{\partial p_y}{p_z\partial\xi} = \frac{ekV}{cp_z}.
\label{eq:S}
\end{equation}
Assuming the optimum case of an imaging optic with $90^\circ$ phase advance between TDS and screen, a criterion for the longitudinal resolution $\Delta\xi$ can then be defined as the angular deflection along the length $\Delta\xi$ being larger than the intrinsic divergence of the unstreaked beam
\begin{equation}
\Delta\xi \ge \frac{\epsilon_y}{\sigma_{y,e}S} = \frac{\epsilon_ycp_z}{\sigma_{y,e}ekV}.
\label{eq:res_long}
\end{equation}
Here, $\epsilon_y$ is the geometric emittance and $\sigma_{y,e}$ is the rms bunch size at its focus, which is at the streaking position. Note that the resolution does not depend on the beam energy but only on the normalized emittance $\epsilon_{ny}$, as the numerator can be rewritten to $\epsilon_y c p_z = \epsilon_{ny} m_e c^2$.

Consider a wakefield driven by a Gaussian laser pulse defined as $a^2=a_0^2\exp{\left(-r^2/(2\sigma_r^2)\right)}\exp{\left(-\zeta^2/(2\sigma_z^2)\right)}$, with $r^2 = x^2 + y^2$, where $a = eA/m_ec^2$ is the normalized laser vector potential and $\sigma_r$ and $\sigma_z$ are the transverse and longitudinal rms width of the laser intensity, respectively. In the linear regime ($a_0^2 \ll 1$), the electric fields far behind the laser are given by \cite{gorbunov1987,esarey2009}
\begin{align}
E_z(r,\zeta) &= E_{z0} \exp{\left(-\frac{r^2}{2\sigma_r^2}\right)}\cos{\left(k_p\zeta\right)}, \label{eq:Ez} \\
E_r(r,\zeta) &= -E_{z0} \frac{r}{k_p\sigma_r^2} \exp{\left(-\frac{r^2}{2\sigma_r^2}\right)}\sin{\left(k_p\zeta\right)}, \label{eq:Er}
\end{align}
with $E_{z0}$ the amplitude of the longitudinal field component and $k_p = \left(ne^2/m_e\epsilon_0c^2\right)^{1/2}=2\pi/\lambda_p$ the plasma wavenumber. Here, $\zeta = z-v_gt$ is the comoving variable ($\zeta=0$ at the laser peak) and $v_g$ the laser group velocity.

\begin{figure}
\centering
\includegraphics[width=9cm]{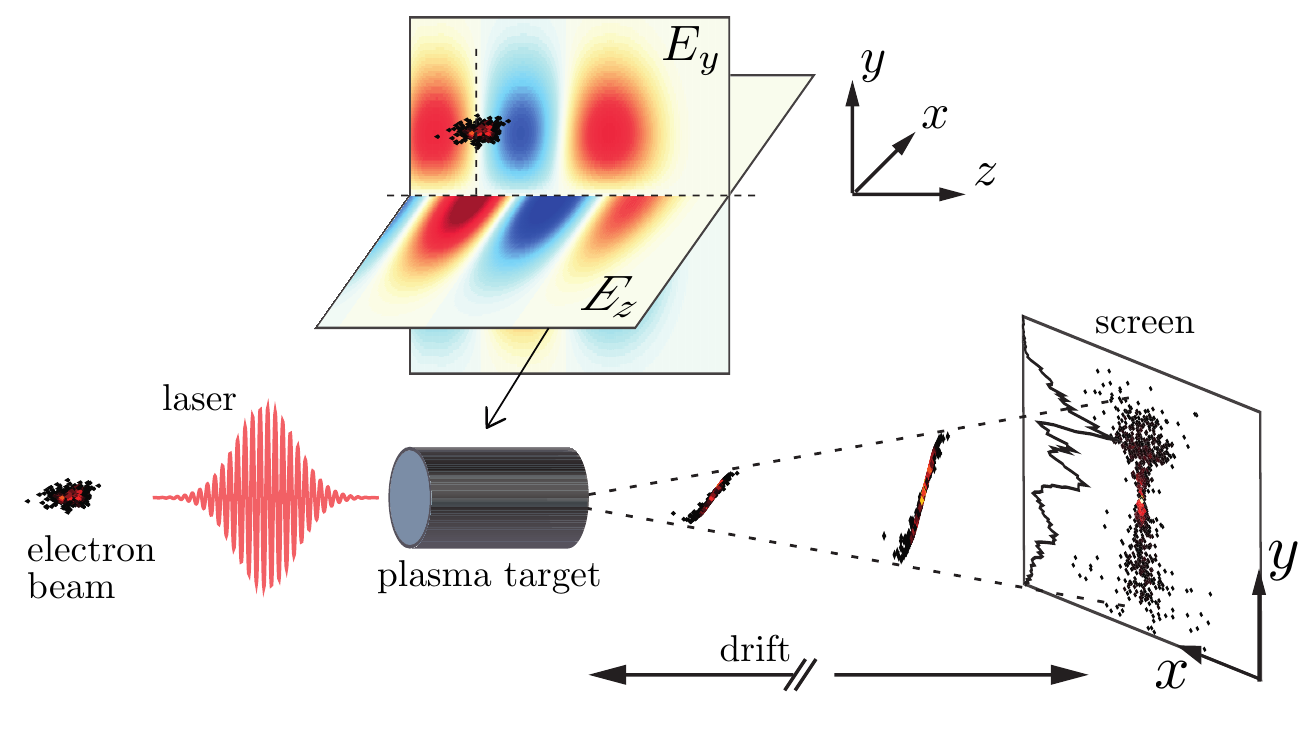}
\caption{Layout of the proposed streaking setup. A short-pulse laser drives a linear wakefield in a plasma target. The electron bunch is situated off-axis at the transverse maximum and longitudinal zero-crossing of the transverse fields. In a drift or imaging optic after the interaction, the imprinted momentum change is translated into a change of transverse position carrying the time information.}
\label{fig:wake_scheme}
\end{figure}

We can realize a TDS-like streaking field using $E_r(r,\zeta)$. Transversally, the bunch should be placed at the maximum of $E_r$, which is at $r = \sigma_r$. If the driver laser and the bunch propagate collinearly, the delay must be chosen as a multiple of $\lambda_p/2$ to position the bunch at the longitudinal zero-crossing of $E_r$.
The setup is illustrated in figure \ref{fig:wake_scheme}.

We will assume that the bunch is displaced by $\sigma_r$ in $y$ with respect to the laser. Assuming a constant wakefield amplitude over a uniform plasma target of length $l$, the transverse momentum change is
\begin{equation}
\Delta p_y = \frac{1}{c}\int eE_y \Big|_{y=\sigma_r}\mathrm dz = \frac{eV}{c}\sin{(k_p\zeta)},
\label{eq:py_wake}
\end{equation}
with
\begin{equation}
V = \frac{E_{z0}l}{k_p\sigma_r}\exp{\left(-\frac{1}{2}\right)}.
\label{eq:V}
\end{equation}
The longitudinal resolution is then given by equation \eqref{eq:res_long}, with $V$ as in eq.\ \eqref{eq:V}.

The transversally quadratic dependence of $E_y$ around $y=\sigma_r$ will result in a spread in the transverse momentum change and a loss of longitudinal resolution, such that
\begin{equation}
\Delta \zeta \gtrsim \sqrt{\frac{5}{2}}\left(\frac{\sigma_{y,e}}{\sigma_r}\right)^2 \vert \zeta \vert.
\label{eq:zeta_quad}
\end{equation}
Therefore $\sigma_{y,e}\ll\sigma_r$ is favorable.

A collinear geometry of laser and electron bunch is directly applicable to external injection experiments. Here, an electron bunch from either an RF-driven electron gun \cite{zeitlerregae2013,assmannsinbad2014}, or from a previous plasma target in a staged setup, is injected into a plasma target for further acceleration. With a simple transverse displacement of target and laser, the plasma target can also act as a bunch diagnostic. Our concept is straightforward to implement in experiments, as only one additional screen is required to monitor the dispersed beam.

The transverse deflecting plasma stage (TDP) can also be operated with an angle between the electron beam and the laser driving the plasma wave. This provides flexibility, especially space for mirrors needed to couple the laser in and out of the electron beam path.

\begin{figure}
\centering
\includegraphics[width=7cm]{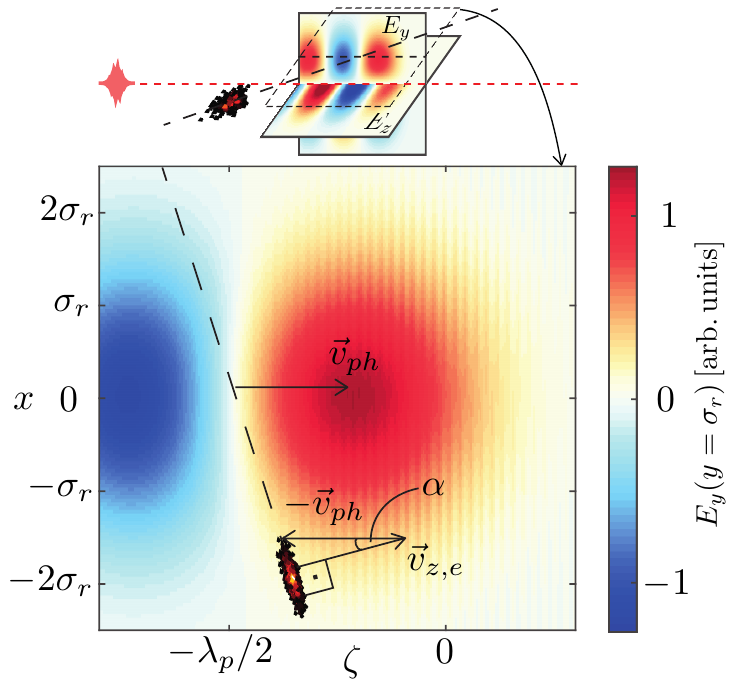}
\caption{Bunch crossing the streaking wakefield at an angle $\alpha$ in the $x$-$z$-plane and at a displacement of $\sigma_r$ in the $y$-direction. The colors represent the amplitude of $E_y$ from a PIC simulation, and the electron beam is Gaussian, for illustration. The shown field slice is indicated by the dashed box on top. Please see text for details.}
\label{fig:setup_angle}
\end{figure}

The geometry of such a setup can be seen in figure \ref{fig:setup_angle}. If the bunch crosses the wakefield in the $x$-$z$-plane an offset needs to be introduced in the $y$ direction to ensure a crossing at the maximum amplitude of $E_y$.
$v_{z,e} = \beta c$ is defined along the direction of movement of the bunch in the lab frame, while $\zeta$ now is rotated by the angle $\alpha$ with respect to $\vec{v}_{z,e}$.
In a frame co-moving with the laser, the bunch moves transversely through the wakefield.
The direction of movement of the bunch in this frame is indicated by the dashed line in fig.\ \ref{fig:setup_angle}.
We require the imprinted streaking signal $\Delta p_y$, which is proportional to the integration of $E_y$ along this line, to be only dependent on the longitudinal bunch coordinate and not on the transverse position in the bunch. Consequently, this line needs to be perpendicular to $\vec{v}_{z,e}$. Otherwise, electrons at the same longitudinal position within the bunch but at different transverse coordinate will experience a different change of momentum, smearing out the streaking trace. The crossing angle $\alpha$ can be found between the bunch velocity $\vec{v}_{z,e}$ and the phase velocity $\vec{v}_{ph}$ of the wakefield at the bunch position. This implies a condition on $v_{ph}$ as $v_{ph} = v_{z,e}/\cos{\alpha}$.

For relativistic beams $v_{z,e} \approx c$, so that a phase velocity $v_{ph}$ larger than $c$ is necessary. This can be achieved by introducing a plasma upramp that leads to the desired phase velocity at the bunch position.
Under the assumption that the laser travels at approximately the speed of light, the plasma ramp can be calculated by solving the differential equation \cite{fubiani2006}
\begin{equation}
\frac{\mathrm{d}n}{\mathrm{d}z} = \left(\frac{c}{v_{ph}} - 1\right) \frac{2n}{\zeta_e}.
\label{eq:dndz}
\end{equation}
The dependence of the bunch position $\zeta_e$ on $z$ can be calculated as $\zeta_e(z) = \zeta_0 + (v_{z,e}/v_{gr}\cos{\alpha} - 1)z$, with the laser group velocity $v_{gr}$.

In the following, we illustrate our concept with simulations. The SINBAD linac \cite{marchetti2015} is a proposed machine currently being designed at DESY for external injection experiments, that has an operation mode for the generation of few- and sub-fs electron bunches. We perform start-to-end simulations of the plasma-streaked SINBAD beam: The acceleration in the linac is simulated with the space charge tracking code \textsc{Astra} \cite{astra}, followed by the plasma interaction simulated with the PIC code \textsc{Warp} \cite{Warp}. The subsequent tracking of the electron bunch along the drift to the screen is again performed with \textsc{Astra}.

We consider a beam of $0.5\,\mathrm{pC}$ charge that is compressed to a few fs bunch length by means of velocity bunching \cite{ferrario2010} and accelerated to \mbox{110 MeV}. The resulting current profile strongly depends on the phase in the accelerating cavities. Here we choose a detuned phase that creates a spiky current profile of \mbox{7.5 fs rms} length to show the capability of our method as a diagnostic or even feedback to tune the phase.
In its focus after the acceleration, where the plasma target is placed, the beam has a transverse size of $\sigma_{x,e} = \sigma_{y,e} = 17\,\mathrm{{\mu}m}$, normalized emittance of $\epsilon_{nx} = \epsilon_{ny} = 0.09\,\mathrm{mm.mrad}$, and $0.17\,\%$ rms energy spread. A large electron beam size is desirable as it reduces the divergence, which in turn determines the background to the imprinted streaking signal.

The laser intensity needs to be chosen low enough so that the wavefront curvature caused by wakefield nonlinearities is negligible. A strong curvature of the wavefronts causes an additional correlation of $E_y$ with $y$ that can smear out the streaking trace. On the other hand, a high laser intensity is desirable as it increases the wakefield amplitude. We choose a laser of $a_0 = 0.3$, $\lambda_l = 0.8\,\mathrm{{\mu}m}$, \mbox{41\,fs} fwhm length and $\sigma_r = 75\,\mathrm{{\mu}m}$ focal spot size (\mbox{3\,J} pulse energy). The laser spot size needs to be significantly larger than the electron beam size to minimize the influence of the quadratic dependence of $E_y$ on $x$ and $y$ at the streaking offset, as indicated by eq.\ \eqref{eq:zeta_quad}. The laser is polarized in $x$ and is focused into the middle of a plasma target of \mbox{3.5\,mm} length and flat top plasma density of $1\cdot10^{18}\,\mathrm{cm^{-3}}$. A high density increases the resolution in two ways, first by increasing $k_p$, and second by increasing the voltage $V$.

The beam is injected externally into the wake at an offset of $y_\text{off} = \sigma_r = 75\,\mathrm{{\mu}m}$ and at a distance behind the driver of $\zeta = -\lambda_p = -34\,\mathrm{{\mu}m}$.

\begin{figure}
\includegraphics{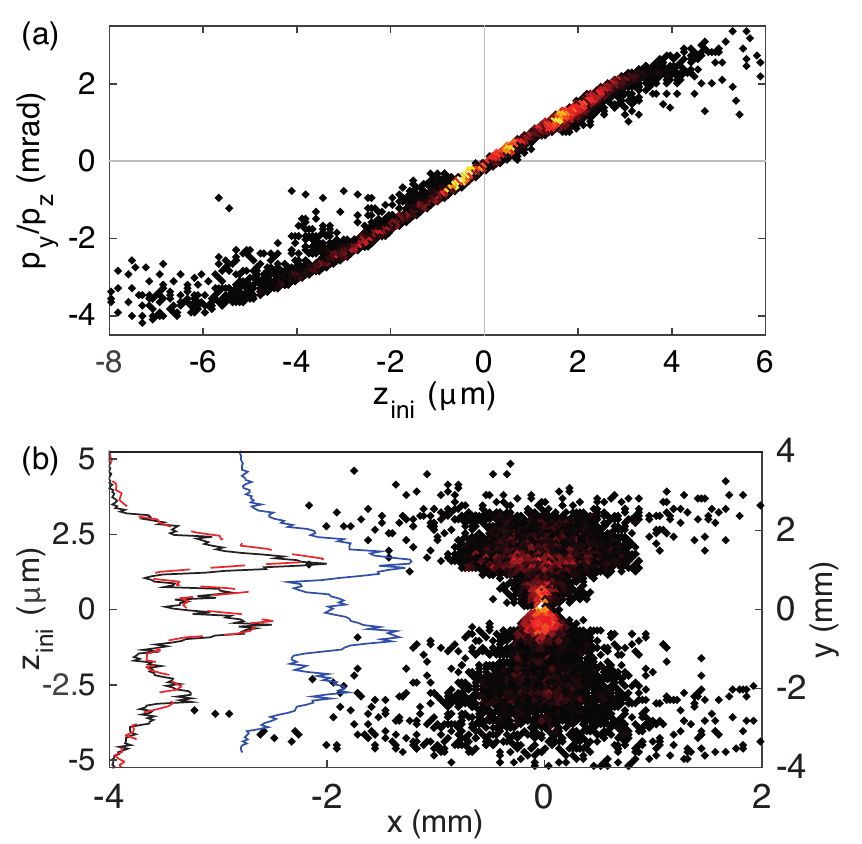}
\caption{Top: Angular deflection $p_y/p_z$ depending on the initial longitudinal position $z_{\text{ini}}$ within the bunch after the interaction with the streaking wakefield. In a drift or imaging optic, this correlation translates to a correlation of $y$ with $z_{\text{ini}}$. Bottom: Screen image at \mbox{1 m} drift after the interaction with the wakefield. The black line is the current profile, reconstructed by binning the screen image to the $y$ axis. The initial current profile is given in dashed red for comparison. For reference, $z_{\text{ini}}$ corresponding to $y$ on screen is given at the left axis. The blue line (displaced for visibility) is the current profile reconstructed from the interaction geometry featuring a $5^\circ$ angle between laser and electron beam path (see text).}
\label{fig:phsp}
\end{figure}

To model the plasma interaction we use \textsc{Warp} in 3D in the boosted frame ($\gamma_{boost} = 10$). The simulation box size is 84 $\mathrm{{\mu}m}$ x 750 $\mathrm{{\mu}m}$ x 750 $\mathrm{{\mu}m}$ with 3150 x 375 x 375 cells and one particle per cell.
During the interaction with the wakefield the bunch accumulates a change of transverse momentum $\Delta p_y$ that is correlated with the longitudinal coordinate, as can be seen in figure \ref{fig:phsp}(a).

At the head and the tail of the bunch the streaking trace is smeared out. This loss of resolution is caused by the quadratic dependence of $E_y$ on $x$ and $y$ (see eq.\ \eqref{eq:zeta_quad}).
Also at head and tail, the curvature caused by the sine-like dependence of $E_y$ on $\zeta$ is visible. Consequently, for streaking longer bunches, a lower plasma density and longer plasma wavelength would be desirable.

For the sake of simplicity we do not use an imaging optics after the plasma. In our specific case, the phase advance introduced by the drift of 1\,m amounts to $50^\circ$, giving rise to a resolution degradation of \mbox{23\,\%} compared to the ideal $90^\circ$. The transverse profile of the beam is then binned to the $y$ axis of a simulated screen, which gives the longitudinal current profile (see figure \ref{fig:phsp}(b)).

The theoretical resolution calculated from equations \eqref{eq:res_long} and \eqref{eq:V} is 96\,as, with $k_p = 1.9\cdot 10^{5}\,\mathrm{m^{-1}}$ and $V = 0.5\,\mathrm{MV}$. This idealized resolution estimate is only valid in the bunch center as it does not account for the second-order correlations of $E_y$ given by eq.\ \eqref{eq:zeta_quad}. The resolution also profits from the low emittance of the electron bunch.

For illustration, another simulation is shown for a geometry where the laser pulse propagates at an angle of $5^\circ$ in the $x$-$z$-plane with respect to the electron beam. Electron and laser beam parameters are the same as above. In contrast, the plasma profile is \mbox{6.5 mm} long and the density rises linearly from $0.26 \cdot 10^{18}\,\mathrm{cm^{-3}}$ to $1.9 \cdot 10^{18}\,\mathrm{cm^{-3}}$. This is a linear approximation to the ideal density profile calculated by numerically solving equation \eqref{eq:dndz}. The electron beam is displaced in the $y$ direction by $y_\text{off} = \sigma_r = 75\,\mathrm{{\mu}m}$.
After the interaction the beam is propagated for \mbox{1 m} as in the collinear geometry. The longitudinal current profile is obtained again by binning to the $y$ axis, and is given in blue in fig.\ \ref{fig:phsp}(b).

Transverse deflecting structures are intrinsically limited either in the achievable temporal or energy spread resolution \cite{behrens2009}. The TDP follows exactly the same reasoning: At the position of maximum $E_y$ field $y = \sigma_r$, the longitudinal electric field of the linear wakefield $E_z$ decays to first order linearly with $y$. This imprints an additional energy spread on the bunch, thereby limiting the slice energy spread resolution achievable by sending the streaked bunch into a spectrometer.
The accumulated energy spread is \cite{behrens2009}
\begin{align}
\frac{\sigma_E}{E} = \frac{eVk_p\sigma_{y,e}}{pc}.
\label{eq:delta_e}
\end{align}
In the example with collinear geometry shown above, the beam exits the plasma with $1.4\,\%$ rms energy spread, which agrees with the $1.4\,\%$ predicted by the eq.\,\eqref{eq:delta_e}.

In our examples, the low bunch charge, which was required to suppress space charge effects at the accelerator gun and to achieve the short bunch length, at the same time also reduces beam loading effects in the plasma. For higher bunch charge, however, beam loading could hamper the resolution by overlaying the transverse field of the laser-driven wake with the beam-driven wakefield. Equation \eqref{eq:py_wake} is then modified by the transverse field of the beam-driven wake to
\begin{equation}
\Delta p_y = \frac{eV}{c}\sin{(k_p\zeta)} + \frac{el}{c}\left(E_{y,b}(r,\zeta)+cB_{x,b}(r,\zeta)\right).
\label{eq:py_mod}
\end{equation}
Following ref.\ \cite{keinigs1987}, the transverse wake of a Gaussian beam for $(k_p\sigma_y)^2 \gg 1$ may be approximated as
\begin{equation}
E_{y,b}+cB_{x,b} = E_0\frac{y}{k_p\sigma_{y,e}^2}\exp{\left(-\frac{r^2}{2\sigma_{y,e}^2}\right)}Z(\zeta),
\end{equation}
where $E_0=m_ec^2k_p/e$ and $Z(\zeta) = \int_\infty^\zeta k_p\hat{n}_b/n\exp{\left(-\zeta^{\prime2}/(2\sigma_{z,e}^2)\right)}\sin{(k_p(\zeta-\zeta^\prime))}\mathrm{d}\zeta^\prime$. For an offset beam with respect to the laser, $y \rightarrow y-\sigma_r$.
According to equ.\ \eqref{eq:py_mod} the mean momentum change is $\langle p_y\rangle = eV/c\sin{(k_p\zeta)}$, and the rms transverse momentum caused by beam loading is
\begin{equation}
\sigma_{py} = \frac{elE_0}{3k_p\sigma_{y,e}c}Z(\zeta).
\end{equation}
The resolution will then be reduced to $\Delta\xi > \sigma_{py}\left(eVk_p/c\right)^{-1}$. Owing to the $\zeta$ dependence in $Z$ the impact of beam loading will increase along the bunch and the bunch head will remain mostly unaffected.
For a longitudinally Gaussian beam, and other beam parameters as in the simulations shown above, the resolution limit at the center of the bunch given by beam loading is small with only $\Delta\xi/c > 66\,\mathrm{as}$. However, for higher charge of \mbox{10 pC} it increases to \mbox{1.3 fs}, which dominates over the theoretical resolution of \mbox{96 as} from eq.\ \eqref{eq:res_long}. For parameter ranges where beam loading is expected to play a role, PIC simulations are necessary to study the influence in depth, since the approximate analytical expressions over-estimate the effect.

Resolution degradation from beam loading could be counteracted by increasing the spot size of both laser and beam while keeping $a_0$ constant, as this decreases the beam density and therefore the amplitude of the beam driven wakefield. $\Delta\xi$ can also be improved by increasing $a_0$ and consequently $V$. Both options call for higher laser pulse energy.

The plasma target length $l$ can be limited by the onset of relativistic self-focusing of the laser or by slippage between the beam and laser. If the laser power $P$ is above the critical power for relativistic self-focusing $P_c\mathrm{(GW)} \approx 17.4 \left(\omega_l/\omega_p\right)^2$, the plasma target length should be significantly shorter than the effective Rayleigh length, i.e., $l^2 \ll z_r^2/(P/P_c-1)$, to avoid strong self-focusing of the laser \cite{esarey2009}. For the parameters above, the Rayleigh length is $z_R = 8.8\,\mathrm{cm}$ and $P/P_c \approx 2$, and this condition is well satisfied.  To neglect slippage effects requires $l \ll \lambda_p^3/\lambda_l^2$, which is also well-satisfied for the example considered.

The longitudinal resolution is also influenced by the plasma density. By tuning the density and consequently the plasma wavelength, the resolution and maximum resolvable bunch length can be flexibly adjusted to suit the expected bunch parameters. The required synchronization between laser and bunch also scales with the density. A jitter of $10\,\%$ of the plasma wavelength at $n=1\cdot 10^{18}\,\mathrm{cm^{-3}}$ requires a synchronization on the \mbox{10 fs} level, which has already been demonstrated at large scale FEL facilities \cite{schulz2015}.
Also other error sources like pointing and positioning jitters could hamper the resolution.

\begin{figure}
\includegraphics{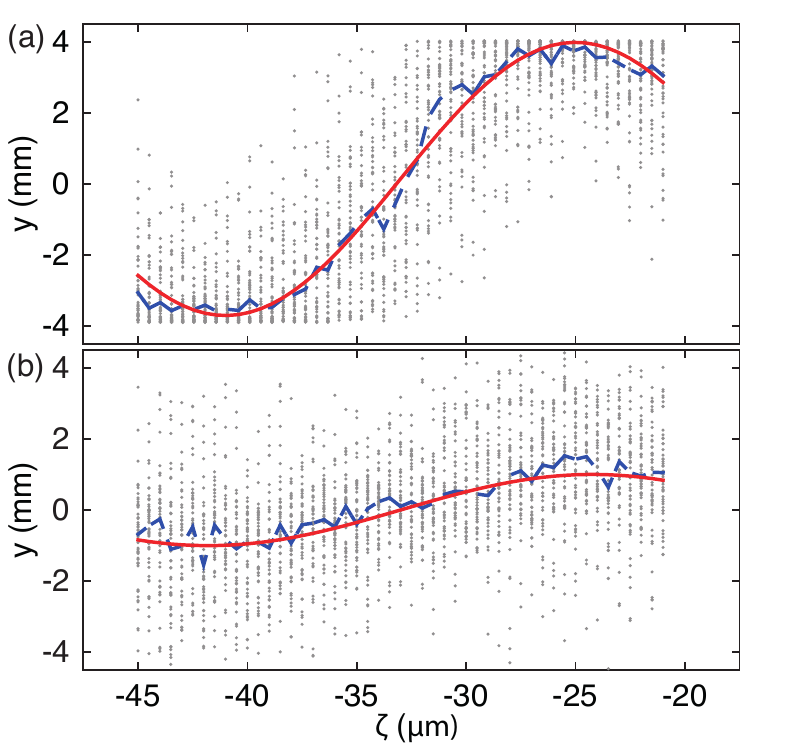}
\caption{\textsc{Astra} simulations of a delay scan including an arrival time jitter of \mbox{10 fs} rms between driver laser and witness bunch (a) or \mbox{10 fs} timing jitter as well as \mbox{500 $\mu$rad} pointing and \mbox{75 $\mu$m} positioning jitter (b). For each delay 50 shots are simulated. Plasma, laser and bunch parameters are like in the PIC simulation from fig.\ \ref{fig:phsp}. The gray dots show the median of each bunch profile in $y$ on the screen. The wavelength is reconstructed from a sine fit (solid red) to the median of all shots at each delay (dashed blue) to (a) $32\,\mathrm{{\mu}m}$ or (b) $35\,\mathrm{{\mu}m}$. The deflection amplitude is obtained from the minimum and maximum deflections to (a) $4.0\,\mathrm{mm}$ or (b) $4.8\,\mathrm{mm}$.}
\label{fig:jitter}
\end{figure}

To demonstrate the feasibility of a TDP deflection calibration in the presence of jitters, simulations are performed with the space charge particle tracking code \textsc{Astra} \cite{astra}. \textsc{Astra} employs the linear wakefield model according to eqs.\ \eqref{eq:Ez} and \eqref{eq:Er}. The laser evolution is defined analytically and slippage effects are included, as well as phase changes due to changes of density. It does not include beam loading. We use the same parameters as in the collinear case discussed above, and study first an arrival time jitter between driver laser and witness bunch of \mbox{10 fs} rms, and also its combination with a positioning and pointing jitter of the electron beam of \mbox{75 $\mu$m} and \mbox{500 $\mu$rad} rms, respectively. This is a conservative estimation of the jitters we expect for external injection experiments in combination with a state-of-the-art \mbox{100 TW} class laser system. The jitter is imprinted on the electron beam instead of the laser, though the laser is likely more prone to these errors. This is an upper estimation of the jitter influence as a pointing jitter of the electron beam leads to an additional steering on top of the signal which is not the case for a pointing jitter of only the laser. To calibrate the TDP, we perform a delay scan around the optimal delay of $\zeta/c = -33\,\mathrm{{\mu}m}/c$ in $0.5\,\mathrm{{\mu}m}/c$ intervals and simulate 50 shots at each delay. The deflection of the bunch centroid at the screen at \mbox{1 m} behind the plasma target can be seen in figure \ref{fig:jitter}. To reduce the influence of the curvature of $E_y(x,y,\zeta)$, which causes an asymmetric bunch profile on the screen, we use the median of the distribution to calculate the centroid rather than its mean. The calibration, i.e.\ $S$, can be calculated from the plasma wavelength and the voltage. The wavelength is reconstructed by fitting a sine function (solid red) to again the median of each 50 shots (dashed blue). For the voltage the maximum positive and negative deflections of all samples is used. The calibration obtained from these jitter-affected simulations differs by \mbox{-6 \%} (or \mbox{+3 \%} for the combination) from the one obtained from a linear fit to the phase space shown in fig.\ \ref{fig:phsp}(a), which is used in fig.\ \ref{fig:phsp}(b).
Experimentally, the calibration can eliminate uncertainties from unknown parameters. For example, both the exact plasma profile as well as the transverse laser profile are not important as long as they are stable, since the imprinted streaking signal is given by the integrated field along the plasma profile. Deviations from the theoretical shape will then be included in the calibration.

In conclusion, we present a new technique that allows to directly diagnose ultrashort electron bunches with a resolution well below 1\,fs. It harnesses the strong transverse fields in a linear plasma wakefield to map the longitudinal bunch profile into a transverse momentum change. Compared to TDS cavities, the increase in resolution is owed to the high field strength and to the short scale of the plasma wavelength.

For optimum resolution, the electron bunch needs to be focused to spot sizes much smaller than the driver laser. The method therefore is well-suited for accelerators intrinsically featuring small beam sizes, such as laser-plasma accelerators or injectors for LPA booster stages.

The technique presented here also drastically shortens the structure length from several meters to millimeters. Such a short interaction length is crucial for applications where the bunch shape changes quickly, which is the case for ultrashort bunches produced by ballistic bunching \cite{Lu2015}. Especially for the diagnostic of laser-plasma accelerated bunches this technique also intrinsically offers good synchronization and therefore jitter tolerance, since parts of the same laser pulse could be used to first generate and then diagnose the bunch.

\begin{acknowledgments}
We would like to thank R. Assmann and M. Titberidze for fruitful discussions.
We gratefully acknowledge funding BMBF Grant 05K13GU5, and the computing time provided on the supercomputers JUROPA and JURECA under project HHH20 and on the PHYSnet cluster of the University of Hamburg.
I.\ Dornmair acknowledges support by the IMPRS UFAST.
Work at LBNL was supported by the Director, Office of Science, Office of High Energy Physics, of the U.S. Department of Energy under Contract No. DE-AC02-05CH11231.
\end{acknowledgments}

\providecommand{\noopsort}[1]{}\providecommand{\singleletter}[1]{#1}%

\end{document}